\documentclass[11pt]{article}
 \usepackage{benstyle}

\usepackage{tikz}
\usetikzlibrary{arrows}
\usepackage{caption}

\usepackage[font=small]{caption}

\title{Compressed sensing with local structure: uniform recovery guarantees for the sparsity in levels class}
\author{Chen Li \\ Courant Institute of Mathematical Sciences \\ New York University
\\ USA
\and 
Ben Adcock \\ Department of Mathematics \\ Simon Fraser University \\ Canada}

\begin{document}
\maketitle

\begin{abstract}
In compressed sensing, it is often desirable to consider signals possessing additional structure beyond sparsity.  One such structured signal model -- which forms the focus of this paper -- is the local sparsity in levels class.  This class has recently found applications in problems such as compressive imaging, multi-sensor acquisition systems and sparse regularization in inverse problems.  In this paper we present uniform recovery guarantees for this class when the measurement matrix corresponds to a subsampled isometry.  We do this by establishing a variant of the standard restricted isometry property for sparse in levels vectors, known as the restricted isometry property in levels.  Interestingly, besides the usual log factors, our uniform recovery guarantees are simpler and less stringent than existing nonuniform recovery guarantees.  For the particular case of discrete Fourier sampling with Haar wavelet sparsity, a corollary of our main theorem yields a new recovery guarantee which improves over the current state-of-the-art.
\end{abstract}

\section{Introduction}\label{s:introduction}
The Restricted Isometry Property (RIP) is an important concept in compressed sensing (CS) \cite{EldarDuarteCSReview,FoucartRauhutCSbook}.  If a matrix $A \in \bbC^{m \times N}$ satisfies an RIP then one is guaranteed stable and robust recovery of all $s$-sparse vectors from the measurements $y = A x$.  Such recovery can be achieved by a number of different algorithms, including convex optimization ($\ell^1$ minimization), greedy or thresholding methods.  Moreover, due to a substantial body of research over the last decade, there are now large classes of matrices known to provably satisfy this property.  These include not only random matrices (e.g.\ of Gaussian or Bernoulli type), but also many structured matrices (randomly-subsampled isometries or random convolutions, for example \cite{KrahmerRauhutStructured}).

Although sparsity is by far the most predominant model in CS, there has also been a significant amount of research focusing on more structured signal models.  See \cite{AHPRBreaking,BourrierEtAlInstance,ModelsCS,DuarteEldarStructuredCS,RauhutWardWeighted,TraonmilinGribonvalRIP} and references therein.  The motivations for this are largely twofold.  First, sparsity has been shown to not be the right model for some existing applications of CS.  Hence to explain its empirical success in such cases one needs to work with other signal models.  Second, if the object to recover is known \textit{a priori} to possess more structure than sparsity alone, i.e.\ it belongs to a structured sparsity class, then it can be highly beneficial  to leverage such additional structure to enhance reconstruction accuracy.  This can be achieved through either the recovery algorithm (so-called \textit{structured recovery} techniques) or, when possible, through the design of the measurement matrix $A$ (so-called \textit{structured sampling} techniques \cite{OptimalSamplingQuest}).

\subsection{Aim of the paper}
Amongst the various models of structured sparsity, the concern of this paper is the so-called \textit{local sparsity in levels} model, introduced in \cite{AHPRBreaking}.  Given a set of $r$ levels (a partition of the indices $\{1,\ldots,N\}$), rather than considering the global sparsity $s$ of a vector $x \in \bbC^{N}$ in this model one considers a vector of $r$ sparsities $\mb{s} = (s_1,\ldots,s_r)$, where $s_k$ is the sparsity of $x$ restricted to the $k^{\rth}$ level.  Although such levels can be chosen arbitrarily, this model is particularly relevant when $x$ is a vector of wavelet coefficients of a signal or image.  In this case, the levels are taken to be wavelet scales and $s_k$ counts the number of nonzero coefficients in the $k^{\rth}$ scale.  The benefit of viewing wavelet coefficients within this model rather than with the standard sparsity model is that it incorporates additional structure inherent to such coefficients.  In particular, the tendency of the local sparsities $s_k$ to decrease as a fraction of the size of the scale as $k$ increases -- a property referred to as \textit{asymptotic sparsity} in \cite{AHPRBreaking,AsymptoticCS}.

In \cite{AHPRBreaking}, the sparsity in levels model was analyzed for the case of a measurement matrix $A \in \bbC^{m \times N}$ arising from randomly subsampling an arbitrary isometry $U \in \bbC^{N \times N}$ according to an appropriate distribution.  The key results proved therein are instances of \textit{nonuniform} recovery guarantees in CS theory.  In other words, they ensure recovery of a fixed vector $x$ for each random draw of the matrix $A$.  Nonuniform CS recovery guarantees do not require an RIP and typically lead to somewhat less stringent measurement conditions by several log factors \cite{Candes_Plan,FoucartRauhutCSbook}.

The purpose of this paper is to complement this work by presenting \textit{uniform} recovery guarantees for such measurement matrices.  We do this by showing that they satisfy an appropriate generalization of the RIP, known as the \textit{Restricted Isometry Property in Levels (RIPL)}.  This property was introduced in \cite{BastounisHansen} (see also \cite{TraonmilinGribonvalRIP}), and like the original RIP is known to be sufficient for stable and robust recovery via $\ell^1$ minimization.

Our proof of the RIPL for subsampled isometries follows a similar route to existing arguments for proving the standard RIP for such matrices \cite{FoucartRauhutCSbook}, albeit with some significant modifications to take into account the new model.  Interestingly, the uniform recovery guarantees we derive -- although less sharp in the usual sense of incurring several additional log factors -- are better than the nonuniform guarantees of \cite{AHPRBreaking} in a number of crucial ways.  In particular, they do not require so-called \textit{relative sparsities} and use a simpler (and fundamentally smaller) notion of \textit{local coherence}.  As in \cite{AHPRBreaking}, our main example will be the case of Fourier sampling with wavelets.  In particular, we prove the uniform version of the nonuniform result proved in \cite{DiscreteFourierHaar} for the discrete Haar wavelet case.

\subsection{Motivations}
The sparsity in levels model has recently proved useful in a number of different contexts.  A first example is the case of Fourier sampling with wavelets as the sparsifying transform -- the typical formulation in some important applications of CS, including Magnetic Resonance Imaging (MRI) \cite{Lustig}, radio interferometry \cite{CSRadioInterferometry}, and others.  Herein it can be shown (via the so-called \textit{flip test} \cite{AsymptoticCS}) that local sparsity in levels, as opposed to global sparsity, is the correct signal model for this problem.  The associated nonuniform recovery guarantees for this model derived in \cite{AHPRBreaking} (see also \cite{BoyerBlockStructured}) explain some key phenomena seen in such applications; notably the resolution dependence of the reconstruction and the influence of the sparsity structure on the best sampling pattern (by this, we mean the set of Fourier frequencies used).  Analysis based on this model provides insight into how to optimize the sampling pattern in these types of applications, as shown practically in \cite{Siemens}.

A second example of the usefulness of this model arises in so-called compressive imaging, with applications including single-pixel and lensless imaging, as well as fluorescence microscopy \cite{OptimalSamplingQuest,AsymptoticCS}.  In such problems one has significant freedom to design the measurement matrix $A$ -- as opposed to the previous setting where Fourier measurements are prescribed by the physical sensor -- with the main constraint being that $A$ should have binary entries.  Since it takes into account the varying sparsities between wavelet scales, the local sparsity in levels model is a more refined model than global sparsity.  Hence, when this model is promoted through a suitable measurement matrix, e.g.\  a subsampled binary Walsh--Hadamard transform, it leads to substantially improved reconstructions over techniques that are based on conventional, sparsity-promoting CS measurements (e.g. random Bernoulli measurements) \cite{AsymptoticCS}.  Note that this is even the case when the standard CS recovery algorithms are replaced by so-called \textit{structured recovery} techniques: namely, techniques which promote the structured sparsity of wavelet coefficients by modifying the recovery algorithm (so-called model-based CS and its generalizations \cite{BaranuikModelCS,HeCarinStructCS,HeCarinTreeCS,TurboAMP}).  Although such approaches often provide some improvement, superior recovery can often be achieved through the \textit{structured sampling} techniques considered in this paper.  We refer to \cite{OptimalSamplingQuest,AsymptoticCS} for further details.

A third, and recent, example of the use of the sparsity in levels model is the case of multi-sensor acquisition, with application to parallel MRI.  This corresponds to physical measurement systems where multiple sensors simultaneously record measurements of a single object.  It transpires that optimal recovery guarantees are possible in this setting for a much broader class of sensing systems with the sparsity in levels model than with the standard sparsity model.  We refer to \cite{Chun&Adcock:16ITW,AdcockChunSubgauss,AdcockChunParallel,Chun&Li&Adcock:16MMSPARSE} for the details.  Finally, we note that the sparsity in levels model also occurs in the problem of compressed sensing with sparsely corrupted measurements \cite{BAEtAlCorruptions,LiCorruptionsConstrApprox}, and has also recently been used in the analysis of sparse techniques for radar \cite{Dorsch2016}.

\section{Preliminaries}
We commence with a review of the relevant background material.

\subsection{Notation}
Throughout $\nm{\cdot}_{p}$ denotes the $\ell^p$-norm on $\bbC^N$.  If $p=2$ then we just write $\nm{\cdot}$ and we denote the corresponding inner product by $\ip{\cdot}{\cdot}$.  We write $A \lesssim B$ to mean there exists a constant independent of all parameters (including, we stress, the number of levels $r$) such that $A \leq C B$, and similarly for $A \gtrsim B$.  

For $1 \leq s \leq N$, recall that a vector $x \in \bbC^{N}$ is $s$-sparse if $\| x \|_0 = | \mathrm{supp}(x) | \leq s$, where $\mathrm{supp}(x) = \{ j : x_j \neq 0 \}$ is the support of $x$.  As is typical, we denote the set of $s$-sparse vectors by $\Sigma_{s}$.  For $x \in \bbC^{N}$, we write
\bes{
\sigma_{s}(x) = \min \left \{ \| x - z \|_{1} : z \in \Sigma_{s} \right \},
}
for the error of the best approximation of $x$ by an $s$-sparse vector, measured in the $\ell^1$ norm.

Let $\Omega \subseteq \{1,\ldots,N\}$.  With slight abuse of notation, we write $P_{\Omega}$ for both the projection $P_{\Omega} \in \bbC^{N \times N}$ with
\bes{
(P_{\Omega}x)_{j} = \left \{ \begin{array}{cc} x_j & j \in \Omega \\ 0 & \mbox{otherwise} \end{array} \right .,\qquad x \in \bbC^N,
}
and the matrix $P_{\Omega} \in \bbC^{|\Omega| \times N}$ with
\bes{
(P_{\Omega}x)_{j} = x_j,\quad j \in \Omega,\qquad x \in \bbC^N.
}
The precise meaning will be clear from the context.  If $\Omega = \{1,\ldots,M\}$ for some $M \in \bbN$ then we merely write $P_{M}$.  Furthermore if $\Omega = \{ M_1+1,\ldots,M_2 \} $ we write $P^{M_1}_{M_2} = P_{\{ M_1+1,\ldots,M_2 \}}$.

\subsection{Compressed sensing with subsampled isometries}
Let $U \in \bbC^{N \times N}$ be an isometry.  A standard CS setup involves subsampling the rows of $U$ uniformly at random.  Specifically, let $\Omega \subseteq \{1,\ldots,N\}$, $|\Omega|=m$ be chosen uniformly at random and form the measurement matrix $A = P_{\Omega} U$.

\defn{[Coherence]
Let $U \in \bbC^{N \times N}$ be an isometry.  The coherence of $U$ is the quantity
\bes{
\mu(U) = \max_{i,j=1,\ldots,N} | U_{ij} |^2.
}
Note that $N^{-1} \leq \mu(U) \leq 1$.
}

Coherence is a key quantity determining the efficiency of the measurement matrix $A = P_{\Omega} U$.  This can be seen from the following nonuniform recovery guarantee:

\thm{[Nonuniform recovery for subsampled isometries and the sparsity model]
\label{t:sparsity_nonuniform}
Let $U \in \bbC^{N \times N}$, $N \geq 2$, be an isometry and $0 < \epsilon < \exp(-1)$.  Fix $x \in \bbC^{N}$ and suppose that $\Omega \subseteq \{1,\ldots,N\}$, $| \Omega | = m$ is chosen uniformly at random.  Let $y = P_{\Omega} U x + e$ where $\| e \| \leq \sqrt{N/m} \eta$ and suppose that $\hat{x}$ is a solution of
\bes{
\min_{z \in \bbC^N} \| z \|_{1}\ \mbox{subject to $\| P_{\Omega} U z - y \| \leq \sqrt{N/m} \eta$}.
}
Then, with probability at least $1-\epsilon$, we have
\bes{
\| x - \hat{x} \| \lesssim  \sigma_{s}(x) +  (1 + L \sqrt{s} ) \eta,
}
where $L = 1 + \frac{\sqrt{\log(\epsilon^{-1})}}{\log(N)}$, provided
\bes{
m \gtrsim s \cdot N \cdot \mu(U) \cdot \log(\epsilon^{-1}) \cdot \log(N).
}
}
This is Theorem 4.4 of \cite{AHPRBreaking} (specialized to the case of one level, and slightly simplified).  Note that one may obtain a rather better error estimate by using the weak RIP approach of \cite{Candes_Plan}.

Theorem \ref{t:sparsity_nonuniform} is termed a nonuniform recovery guarantee since it ensures the recovery of a fixed vector $x$ for any one draw of $\Omega$.  To derive a uniform guarantee, the standard approach is to use the restricted isometry property:

\defn{[Restricted Isometry Property]
Let $1 \leq s \leq N$.  The $s^{\rth}$ Restricted Isometry Constant (RIC) $\delta_s$ of a matrix $A \in \bbC^{m \times N}$ is the smallest $\delta \geq 0$ such that
\bes{
(1-\delta) \| x \|^2 \leq \| A x \|^2 \leq (1+\delta) \| x \|^2,\quad \forall x \in \Sigma_{s}.
}
If $0 < \delta_s < 1$ we say that A has the Restricted Isometry Property (RIP) of order $s$.
}

\thm{[Stable and robust recovery with the RIP]
\label{t:RIP_recov}
Suppose that $A \in \bbC^{m \times N}$ satisfies the RIP of order $2s$ with constant $\delta_{2s} < 4/\sqrt{41}$.  Let $x \in \bbC^{N}$ and $y = A x + e$ where $\| e \| \leq \eta$.  Then for any minimizer $\hat{x} \in \bbC^N$ of
\bes{
\min_{z \in \bbC^N} \| z \|_{1}\ \mbox{subject to $\| P_{\Omega} U z - y \| \leq  \eta$},
}
we have
\eas{
\| \hat{x} - x \|_{1} \lesssim \sigma_{s}(x) + \sqrt{s} \eta,\qquad \| \hat{x} - x \| \lesssim \frac{1}{\sqrt{s}} \sigma_{s}(x) + \eta.
}
}

This result is Theorem 6.12 of \cite{FoucartRauhutCSbook}.  We note in passing that the inequality can be improved to the sharp value $\delta_{2s} < 1/\sqrt{2}$ \cite{CaiZhangRIP}.  In view of this result, to prove uniform recovery guarantees it suffices establish an RIP.  The following is a typical result for subsampled isometries:

\thm{[Subsampled isometries and the RIP]
\label{t:iso_RIP}
Let $U \in \bbC^{N \times N}$ be an isometry and $0 < \epsilon,\delta < 1$.  Let $t_1,\ldots,t_m$ be chosen uniformly and independently from the set $\{1,\ldots,N\}$ and set $\Omega = \{ t_1,\ldots,t_m \}$.  If
\bes{
m \gtrsim \delta^{-2} \cdot s \cdot N \cdot \mu(U) \cdot \left (  \log(2m) \log(2N) \log^2(2s) + \log( \epsilon^{-1}) \right ),
}
then, with probability at least $1-\epsilon$ the matrix $A = P_{\Omega} U \in \bbC^{m \times N}$ satisfies the RIP of order $s$ with constant $\delta_{s} \leq \delta$.
}

This result is equivalent to Theorem 12.32 of \cite{FoucartRauhutCSbook} (specialized to the case of isometries), and is in fact a consequence of our main result (Theorem \ref{t:RIPlevels}) when the number of levels $r$ is equal to one.  Note that the construction of the set $\Omega$ is slightly different to that of Theorem \ref{t:sparsity_nonuniform}.  This construction transpires to be easier to analyze in practice.  We refer to \cite[Chpt.\ 12]{FoucartRauhutCSbook} for details on different drawing models.  We also remark that it is possible to improve the log factors in the main estimate somewhat.  See \S \ref{s:proof} for some further discussion on this topic.

\subsection{Compressed sensing with local structure: the framework of \cite{AHPRBreaking}}\label{ss:CSlocal}

The framework of \cite{AHPRBreaking} introduced a new structured sparsity model, wherein a vector $x$ is allowed to have different sparsities in separate levels.  The precise definition is as follows:

\defn{[Sparsity in levels]
For $r \in \bbN$, let $\mb{M} = (M_1,\ldots,M_r)$, where $1 \leq M_1 < \ldots < M_{r} = N$, and $\mathbf{s} = (s_1,\ldots,s_r)$, where $s_{k} \leq M_k - M_{k-1}$ for $k=1,\ldots,r$ and $M_0 = 0$.  A vector $x \in \bbC^N$ is $(\mb{s},\mb{M})$-sparse in levels if
\bes{
\left | \mathrm{supp}(x) \cap \{ M_{k-1}+1,\ldots,M_k \} \right | \leq s_k,\quad k=1,\ldots,r.
}
We denote the set of $(\mb{s},\mb{M})$-sparse vectors by $\Sigma_{\mb{s},\mb{M}}$.
}

Note that the vector $\mb{M}$ describes the $r$ \textit{sparsity} levels, and the vector $\mb{s}$ enumerates the \textit{local sparsities} within them.  In an analogous manner to the case of standard sparsity, we write
\be{
\label{best_sM}
\sigma_{\mb{s},\mb{M}}(x) = \left \{ \| x - z \|_1 : z \in \Sigma_{\mb{s},\mb{M}} \right \},
}
for the error of the best approximation by an $(\mb{s},\mb{M})$-sparse vector.

A new approach to subsampling isometries in order to promote the sparsity in levels signal model was also introduced in \cite{AHPRBreaking},  as was a notion of local coherence that refines the usual global coherence quantity in accordance with this model.  These are defined as follows:

\defn{[Multilevel random subsampling]
For $r \in \bbN$, let $\mb{N} = (N_1,\ldots,N_r)$, where $1 \leq N_1 < \ldots < N_{r} = N$, and $\mb{m} = (m_1,\ldots,m_r)$, where $m_{k} \leq N_k - N_{k-1}$ for $k=1,\ldots,r$, and $N_0 = 0$.  For each $k=1,\ldots,r$, let $t_{k,1},\ldots,t_{k,m_k}$ be chosen uniformly and independently from the set $\{N_{k-1}+1,\ldots,N_k\}$, and set $\Omega_{k} = \{ t_{k,1} , \ldots, t_{k,m_k} \}$.  If $\Omega = \Omega_{\mb{N},\mb{m}} = \Omega_1 \cup \cdots \cup \Omega_r$ we refer to $\Omega$ as an $(\mb{N},\mb{m})$-multilevel subsampling scheme.
}
To distinguish them from the sparsity levels, indexed by $\mb{M}$, we refer to the levels indexed by $\mb{N}$ as \textit{sampling} levels.  Note that, similarly to as mentioned above,  the probability model used in \cite{AHPRBreaking} was slightly different (namely, the set $\Omega_k$ was chosen uniformly at random of size $m_k$).

\defn{[Local coherence in levels]
\label{d:loc_coh}
Let $\mb{N} = (N_1,\ldots,N_r)$ and $\mb{M} = (M_1,\ldots,M_r)$ denote sampling and sparsity levels respectively.  The $(k,l)^{\rth}$ local coherence of an isometry $U \in \bbC^{N \times N}$ is
\bes{
\mu_{k,l} = \mu_{k,l}(\mb{N},\mb{M})= \max\left \{  | U_{ij} |^2 : i = N_{k-1}+1,\ldots,N_{k},\ j=
M_{l-1}+1,\ldots,M_l \right \}.
}
}
In \cite{AHPRBreaking}, the local coherences are defined rather differently.  Specifically, one sets
\be{
\label{nonunif_local}
\tilde{\mu}_{k,l} = \tilde{\mu}_{k,l}(\mb{N},\mb{M}) = \max_{t=1,\ldots,r} \sqrt{\mu_{k,l} \mu_{k,t}}.
}
Note that $\tilde{\mu}_{k,l} \geq \mu_{k,l}$.  It transpires that for nonuniform recovery guarantees one needs to consider \R{nonunif_local}, whereas, as we shall show in our main result later, in the uniform case one can work with the simpler and strictly smaller quantities $\mu_{k,l}$.

As discussed in \cite{AHPRBreaking}, for applications of interest, e.g.\ Fourier sampling with wavelet sparsity, the global coherence of the corresponding matrix $U$ is often high, meaning that it is impossible to recover $s$-sparse vectors by subsampling uniformly at random.  However, the behaviour of the local coherence is often such that one can get near-optimal recovery guarantees for the sparsity in levels models when subsampling with a multilevel sampling scheme.   To confirm this, the following nonuniform recovery guarantee was proved in \cite{AHPRBreaking}:

\thm{[Nonuniform recovery for subsampled isometries and the sparsity in levels model]
\label{t:nonunifom_sparsitylevels}
Let  $U \in \bbC^{N \times N}$ be an isometry, $N \geq 2$, and $0 < \epsilon < \exp(-1)$.  Fix $x \in \bbC^{N}$ and suppose that $\Omega = \Omega_{\mb{N},\mb{m}}$ is an $(\mb{N},\mb{m})$-multilevel subsampling scheme.  Let $y = P_{\Omega} U x + e$, where $\| e \| \leq \sqrt{K} \eta$ and $K = \max_{k=1,\ldots,r} \left \{ \frac{N_k-N_{k-1}}{m_k} \right \}$ and suppose that $\hat{x}$ is a solution of
\bes{
\min_{z \in \bbC^N} \| z \|_{1}\ \mbox{subject to $\| P_{\Omega} U z - y \| \leq \sqrt{K} \eta$}.
}
Then with probability at least $1-\epsilon$ we have
\bes{
\| x - \hat{x} \| \lesssim  \sigma_{\mb{s},\mb{M}}(x) +  (1 + L \sqrt{s} ) \eta,
}
where $L = 1 + \frac{\sqrt{\log(\epsilon^{-1})}}{\log(N)}$ and $s = s_1+\ldots + s_r$, provided
\be{
\label{nonunif_mk_1}
m_k \gtrsim (N_k-N_{k-1}) \cdot \left ( \sum^{r}_{l=1} \tilde{\mu}_{k,l} \cdot s_l \right ) \cdot \log(s \epsilon^{-1}) \cdot \log(N),\qquad k=1,\ldots,r,
}
and $m_k \gtrsim \hat{m}_k \cdot \log( s \epsilon^{-1}) \cdot \log(N)$, where the $\hat{m}_k$ are such that
\be{
\label{nonunif_mk_2}
1 \gtrsim \sum^{r}_{k=1} \left ( \frac{N_k-N_{k-1}}{\hat{m}_k} - 1 \right ) \cdot \tilde{\mu}_{k,l}\cdot S_k,\qquad l=1,\ldots,r.
}
Here $S_k = S_k(\mb{N},\mb{M},\mb{s})$ is as in Definition \ref{d:rel_sparsity} and $\tilde{\mu}_{k,l} = \tilde{\mu}_{k,l}(\mb{N},\mb{M})$ is as in \R{nonunif_local}.
}

This theorem relies on the notion of relative sparsities, defined as follows:

\defn{[Relative sparsity]
\label{d:rel_sparsity}
Let $U \in \bbC^{N \times N}$ be an isometry.  Let $\mb{N} = (N_1,\ldots,N_r)$, $\mb{M} = (M_1,\ldots,M_r)$ and $\mb{s} = (s_1,\ldots,s_r)$ define sampling levels, sparsity levels and local sparsities respectively.  Then for $k=1,\ldots,r$ the $k^{\rth}$ relative sparsity $S_k = S_k(\mb{N},\mb{M},\mb{s})$ is given by
\bes{
S_k(\mb{N},\mb{M},\mb{s}) = \max \left \{ \| P^{N_{k-1}}_{N_k} U z\|^2 : \| z \|_{\infty} \leq 1, | \supp(z) \cap \{ M_{l-1}+1,\ldots,M_l \} | \leq s_l,\ l=1,\ldots,r  \right \}.
}
}

\subsection{The case of 1D Fourier sampling with Haar wavelet sparsity}
\label{ss:FourHaar}
As mentioned, an important instance of this general framework is that of Fourier sampling with wavelets as the sparsifying transform.  To illustrate this application, we consider the discrete, one-dimensional setting with the Haar wavelet basis.  This example was discussed in detail in \cite{DiscreteFourierHaar}.

Let $N = 2^r$ for some $r \in \bbN$.  If $x = \{ x_i \}^{N}_{i=1} \in \bbC^N$ we define the Fourier transform as
\bes{
\cF x(\omega) = \frac{1}{\sqrt{N}} \sum^{N}_{j=1} x_{j} \exp(2 \pi \I (j-1) \omega / N ),\qquad \omega \in \bbR,
}
and write $F \in \bbC^{N \times N}$ for the corresponding unitary matrix of this transform, so that $F x = \{ \cF x(\omega) \}^{N/2}_{\omega=-N/2+1}$.  Let $\Phi \in \bbC^{N \times N}$ be the unitary matrix whose columns are the orthonormal Haar basis vectors on $\bbC^N$, and suppose that $x \in \bbC^N$ is a signal with approximately sparse representation in this basis.  We now seek to recover $x$ from $m$ noisy Fourier samples by solving the $\ell^1$ minimization problem
\bes{
\min_{z \in \bbC^N} \| \Phi^* z \|_{1}\ \mbox{subject to $\| P_{\Omega} F z - y \| \leq \eta$}.
}
Here $y = P_{\Omega} F x + e$ and $e$ satisfies $\nm{e} \leq \eta$.  Equivalently, if $x = \Phi w$ where $w \in \bbC^N$ is the vector of Haar wavelet coefficients of $x$, then we solve
\bes{
\min_{z \in \bbC^N} \| z \|_{1}\ \mbox{subject to $\| P_{\Omega} U z - y \| \leq \eta$},
}
where $U = F \Phi$ is an isometry.  Note that if $\hat{w}$ is a minimizer of this problem then $w\approx \hat{w} $ provided $w$ is sufficiently sparse, i.e. the signal $x$ is recovered via $x \approx \hat{x} =  \Phi \hat{w}$.

\begin{figure}[t]
\centering
\begin{tikzpicture}[>=stealth]
\filldraw[fill=black!20,draw=black!50!black, thick] 
(-0.5,0) -- (0.5,0) -- (0.5,4) -- (-0.5,4) -- (-0.5,0);
\filldraw[fill=blue!20,draw=blue!50!black, thick] 
(0.5,0) -- (1.5,0) -- (1.5,4) -- (0.5,4) -- (0.5,0);
\filldraw[fill=blue!20,draw=blue!50!black, thick] 
(-0.5,0) -- (-1.5,0) -- (-1.5,4) -- (-0.5,4) -- (-0.5,0);
\filldraw[fill=red!20,draw=red!50!black, thick] 
(1.5,0) -- (3.5,0) -- (3.5,2) -- (1.5,2) -- (1.5,0);
\filldraw[fill=red!20,draw=red!50!black, thick] 
(-1.5,0) -- (-3.5,0) -- (-3.5,2) -- (-1.5,2) -- (-1.5,0);
\filldraw[fill=green!20,draw=green!50!black, thick] 
(3.5,0) -- (7.5,0) -- (7.5,1) -- (3.5,1) -- (3.5,0);
\filldraw[fill=green!20,draw=green!50!black, thick] 
(-3.5,0) -- (-7.5,0) -- (-7.5,1) -- (-3.5,1) -- (-3.5,0);

\draw (0,2) node [fill=none] {\small $W_1$}  -- (0,2) ;
\draw (1,2) node [fill=none] {\small $W_2$}  -- (1,2) ;
\draw (-1,2) node [fill=none] {\small $W_2$}  -- (-1,2) ;
\draw (2.5,1) node [fill=none] {\small $W_3$}  -- (2.5,1) ;
\draw (-2.5,1) node [fill=none] {\small $W_3$}  -- (-2.5,1) ;
\draw (5.5,0.5) node [fill=none] {\small $W_4$}  -- (5.5,0.5) ;
\draw (-5.5,0.5) node [fill=none] {\small $W_4$}  -- (-5.5,0.5) ;

\draw (0,4.25) node [fill=none] {\small $100\%$} -- (0,4.25);
\draw (1,4.25) node [fill=none] {\small $100\%$} -- (1,4.25);
\draw (-1,4.25) node [fill=none] {\small $100\%$} -- (-1,4.25);
\draw (2.5,2.25) node [fill=none] {\small $50\%$} -- (2.5,2.25);
\draw (-2.5,2.25) node [fill=none] {\small $50\%$} -- (-2.5,2.25);
\draw (5.5,1.25) node [fill=none] {\small $25\%$} -- (5.5,1.25);
\draw (-5.5,1.25) node [fill=none] {\small $25\%$} -- (-5.5,1.25);
\end{tikzpicture}
\caption{Dyadic bands $W_{k}$ and typical sampling ratios $m_k / (N_k - N_{k-1})$ for the multilevel sampling pattern in the case of Fourier sampling with Haar wavelets.  The first two bands $W_1$ and $W_2$ are fully sampled, and the remainder are subsampled.  As is typical in practice, more subsampling is used at higher frequencies to capture the asymptotic sparsity of the wavelet coefficients \cite{AHPRBreaking,AsymptoticCS}.  }
\label{f:fig}
\end{figure}
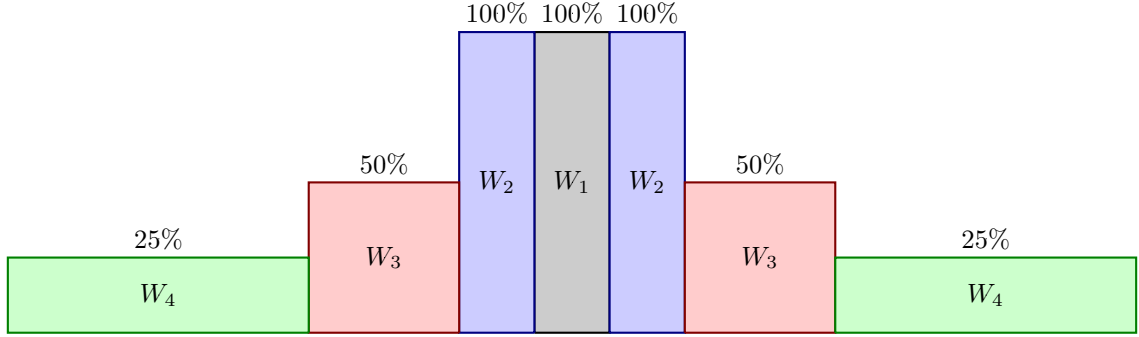

It is well-known that the above matrix $U$ is coherent, i.e.\ $\mu(U)=1$.  Hence choosing $\Omega$ uniformly at random gives poor reconstructions in practice.  Fortunately, near-optimal recovery is possible using the setup of \S \ref{ss:CSlocal}.  To see this, we first need to define the levels $\mb{N}$ and $\mb{M}$.  As in \cite{DiscreteFourierHaar}, we order the Haar basis so that the first level $\{ M_{0}+1,M_1 \} = \{1,2\}$ contains the coefficients of the scaling function and the mother wavelet, and subsequent levels $\{ M_{k-1}+1,\ldots,M_k \} = \{ 2^{k-1}+1,\ldots, 2^k \}$ contain the coefficients of the wavelets at scale $k-1$.  Note that this gives the following:
\be{
\label{FourWav_M}
M_0=0,\  M_{k} = 2^{k},\qquad k=1,\ldots,r.
}
We next introduce the sampling levels.  Following an idea of \cite{Candes_Romberg}, we consider dyadic bands in frequency space.  Specifically, we let $W_{1} = \{ 0,1\}$ and
\bes{
W_{k+1} = \{ -2^{k}+1,\ldots,-2^{k-1} \} \cup \{ 2^{k-1}+1,\ldots,2^k \},\quad k=1,\ldots,r-1.
}
We then choose the sampling levels $\mb{N}$ so that (after reordering the rows of $F$ suitably) the $k^{\rth}$ level $\{ N_{k-1}+1,\ldots,N_k\}$ corresponds to the frequencies in $W_k$.  Note that this gives
\be{
\label{FourWav_N}
N_0=0,\  N_{k} = 2^{k},\qquad k=1,\ldots,r.
}
Figure \ref{f:fig} gives an illustration of a typical sampling pattern using these levels.

With this in hand, one now has the following result:

\cor{[Nonuniform recovery for the 1D Fourier/Haar wavelet system]
\label{c:nonunif_FourHaar}
Let $N=2^r$ for some $r \geq 1$, $0 < \epsilon < \exp(-1)$ and $x \in \bbC^{N}$.  For each $k=1,\ldots,r$ suppose that $m_k$ Fourier samples are drawn uniformly at random from the band $W_k$, where
\be{
\label{nonunif_mk_Haar}
m_k \gtrsim \left ( s_k + \sum^{r}_{\substack{l =1 \\ l \neq k}} 2^{-|k-l|/2} s_l \right ) \cdot \log(s \epsilon^{-1}) \cdot \log(N),
}
for some $s_k \leq M_k - M_{k-1}$, $k=1,\ldots,r$, the $M_k$ are as in \R{FourWav_M} and $s = s_1+\ldots+s_r$.  Let $K = \max_{k=1,\ldots,r} \left \{ \frac{N_k-N_{k-1}}{m_k} \right \}$, where the $N_k$ are as in \R{FourWav_N}, and suppose that $y = P_{\Omega} F x + e$ with $\| e \| \leq \sqrt{K} \eta$ for some $\eta \geq 0$.  Then with probability at least $1-\epsilon$ any minimizer $\hat{x}$ of
\bes{
\min_{z \in \bbC^N} \| \Phi^* z \|_{1}\ \mbox{subject to $\| P_{\Omega} F z - y \| \leq \sqrt{K} \eta$}.
}
satisfies
\bes{
\| x - \hat{x} \| \lesssim \sigma_{\mb{s},\mb{M}}(\Phi^* x) + (1+L \sqrt{s}) \eta,
}
where $L = 1 + \frac{\sqrt{\log(\epsilon^{-1})}}{\log(N)}$ and $\sigma_{\mb{s},\mb{M}}$ is as in \R{best_sM}.
}

This is Theorem 2 of \cite{DiscreteFourierHaar}.  The result follows from Theorem \ref{t:nonunifom_sparsitylevels} after determining appropriate estimates for the local coherences $\tilde{\mu}_{k,l}(\mb{N},\mb{M})$ and relative sparsities $S_k(\mb{N},\mb{M},\mb{s})$ for the Fourier/wavelets matrix.  These are as follows:
\bes{
\tilde{\mu}_{k,l}(\mb{N},\mb{M}) \lesssim 2^{-k} 2^{-|k-l|/2},\quad k,l=1,\ldots,r,
}
and
\bes{
S_k(\mb{N},\mb{M},\mb{s}) \lesssim \sum^{r}_{l=1} 2^{-|k-l|/2} s_l.
}
See \cite[eqn.\ (12)]{DiscreteFourierHaar} and \cite[eqn.\ (17)]{DiscreteFourierHaar} respectively.

The main estimate \R{nonunif_mk_Haar} states that the number of measurements $m_k$ required in the $k^{\rth}$ sampling band is proportional to the number of nonzero coefficients in the corresponding wavelet scale plus exponentially-decaying coupling terms.  The appearance of these terms is due to \textit{interference} between wavelet scales \cite{DiscreteFourierHaar}.  Specifically, the matrix $U$, although dominated by its diagonal blocks, is not exactly block-diagonal.  Hence in the $k^{\rth}$ sampling level there are contributions from not only the corresponding $k^{\rth}$ sparsity level, but also all other levels.  Note that the optimal recovery guarantee in this setup would be $m_{k} \gtrsim s_k$ (up to log factors), and this would have resulted had $U$ been exactly block diagonal.

\subsection{The restricted isometry property in levels}
\label{ss:RIPL}
This paper is devoted to proving uniform recovery guarantees for the sparsity in levels model combined with the framework of  local coherence in levels and multilevel random subsampling.  Much as in standard, sparsity-based CS, this will be done by establishing the following generalized RIP condition:

\begin{definition}[RIP in levels]
Let $\mb{M} = (M_1,\ldots,M_r)$ be sparsity levels and $\mb{s} = (s_1,\ldots,s_r)$ be local sparsities.  The $\mb{s}^{\rth}$ restricted isometry constant in levels (RICL) $\delta_{\mb{s},\mb{M}}$ of a matrix $A \in \bbC^{m \times N}$ is the smallest $\delta \geq 0$ such that
\bes{
(1-\delta) \| x \|^2 \leq \| A x \|^2 \leq (1+\delta) \| x \|^2,\quad \forall x \in \Sigma_{\mb{s},\mb{M}}.
}
If $0 < \delta_{\mb{s},\mb{M}} < 1$ we say that the matrix $A$ satisfies the Restricted Isometry Property in levels (RIPL) of order $(\mb{s},\mb{M})$.
\end{definition}

The RIPL was first introduced in \cite{BastounisHansen}.  Analogous to Theorem \ref{t:RIP_recov}, it implies stable and robust recovery of $\ell^1$ minimization (see \cite[Thm.\ 4.4]{BastounisHansen}):

\thm{[Stable and robust recovery with the RIPL]
\label{t:RIPLstabrobust}
Suppose that $A \in \bbC^{m \times N}$ satisfies the RIPL of order $(2\mb{s},\mb{M})$ with constant
\be{
\label{RIPL_suffic}
\delta_{2\mb{s},\mb{M}} < \frac{1}{\sqrt{r \left ( \sqrt{\rho}+1/4 \right )^2 + 1 }},
}
where
\be{
\label{sparsity_ratio}
\rho = \rho_{\mb{s},\mb{M}} = \max_{k,l=1,\ldots,r} \{ s_k / s_l \}.
}
Let $x \in \bbC^{N}$ and $y = A x + e$ where $\| e \| \leq \eta$.  Then, for any minimizer $\hat{x} \in \bbC^N$ of
\be{
\label{RIPL_recovery}
\min_{z \in \bbC^N} \| z \|_{1}\ \mbox{subject to $\| A z - y \| \leq  \eta$},
}
we have
\eas{
\| \hat{x} - x \|_{1} \lesssim \sigma_{\mb{s},\mb{M}}(x) + \sqrt{s} \eta,
}
and
\be{
\label{Alex}
\| \hat{x} - x \| \lesssim \left ( 1 + (r \rho )^{1/4} \right )\frac{\sigma_{\mb{s},\mb{M}}(x) }{\sqrt{s}}  + (1 + (r \rho )^{1/4} )\eta ,
}
where $s = s_1+\ldots+s_r$.
}

Note that this result implies the standard RIP result Theorem \ref{t:RIP_recov}, which corresponds to the case $r=1$.  As discussed in \cite{BastounisHansen}, the asymptotic scaling of the right-hand side of \R{RIPL_suffic} with both $r$ and $\rho$ is optimal.  The same is true of \R{Alex}.  The appearance of the sparsity ratio $\rho$ in these estimates is particularly unfortunate.  Fortunately, as shown recently in \cite{TraonmilinGribonvalRIP}, it can be removed by replacing the $\ell^1$ minimization term in the optimization problem \R{RIPL_recovery} by a weighted $\ell^1$ minimization functional of the form
\be{
\label{Remiregularizer}
\sum^{r}_{k=1} \frac{1}{ \sqrt{s_k}} \nm{P^{M_{k-1}}_{M_k} z }_{1}.
}
In other words, the entries of $z$ in the $k^{\rth}$ sparsity level are weighted by the corresponding local sparsity $s_k$ in that level.

\section{Main results}
Our main results identify sufficient conditions under which an isometry, subsampled according to a multilevel subsampling scheme and scaled in a way below, satisfies the RIPL.  The scaling is as follows.  Given an isometry $U \in \bbC^{N \times N}$ and a multilevel subsampling scheme $\Omega = \Omega_{\mb{N},\mb{m}} = \Omega_1 \cup \cdots \cup \Omega_r$, we define the matrix
\be{
\label{A_multilevel}
A = \left [ \begin{array}{c} 1/\sqrt{p_1} P_{\Omega_1} U \\ 1/\sqrt{p_2} P_{\Omega_2} U \\ \vdots \\ 1/\sqrt{p_r} P_{\Omega_r} U \end{array} \right ] \in \bbC^{m\times N},\qquad m = m_1+\ldots + m_r,
}
where $p_k = \frac{m_k}{N_k-N_{k-1}}$ for $k=1,\ldots,r$.  Our first result is now as follows:

\thm{
[Subsampled isometries and the RIPL I]
\label{t:RIPlevels}
Let $U \in \bbC^{N \times N}$ be an isometry, $r \in \bbN$ and $0 < \epsilon,\delta < 1$.  Let $\Omega = \Omega_{\mb{N},\mb{m}}$ be an $(\mb{N},\mb{m})$-multilevel subsampling scheme, and $\mb{M}$ and $\mb{s}$ be sparsity levels and local sparsities respectively.  Suppose that
\be{
\label{mk_RIPL}
m_k \gtrsim \delta^{-2} \cdot (N_k - N_{k-1}) \cdot \left ( \sum^{r}_{l=1} \mu_{k,l} \cdot s_l \right ) \cdot \left ( r \log(2m) \log(2N) \log^2(2s) + \log(\epsilon^{-1}) \right ),
}
for $k=1,\ldots,r$, where $m = m_1+\ldots + m_r$.  Then with probability at least $1-\epsilon$, the matrix \R{A_multilevel} satisfies the RIPL of order $(\mb{s},\mb{M})$ with constant $\delta_{\mb{s},\mb{M}} \leq \delta$.
}

The proof of this theorem is given in \S \ref{s:proof}.  It is informative to compare the main condition \R{mk_RIPL} to those of the nonuniform recovery guarantee of Theorem \ref{t:nonunifom_sparsitylevels}.  First, note that the relative sparsities $S_k$, which are required in the nonuniform case through \R{nonunif_mk_2}, are no longer needed in the uniform condition \R{mk_RIPL}.  This greatly simplifies matters in practice, since the relative sparsities require a nontrivial estimation that depends crucially on the choice of $U$ (see \cite{DiscreteFourierHaar} for the Fourier/Haar wavelet case).  Second, besides the log factors, the condition \R{mk_RIPL} is smaller than the corresponding condition \R{nonunif_mk_1} since it involves the local coherences $\mu_{k,l}$ as opposed to the strictly larger quantities $\tilde{\mu}_{k,l}$ defined in \R{nonunif_local}.  This observation will be of particular relevance in the Fourier/Haar wavelet case, as it will lead to a better estimate than the corresponding nonuniform recovery guarantee (Corollary \ref{c:nonunif_FourHaar}).  We will discuss this further in \S \ref{ss:unif_FourHaar}.  On the other hand, due to Theorem \ref{t:RIPLstabrobust} the uniform recovery guarantee also requires $\delta^{-2} \gtrsim r \rho$ (or $\delta^{-2} \gtrsim r$ if the weighted regularizer \R{Remiregularizer} is used).  Thus, \R{mk_RIPL} effectively implies a quadratic scaling with the number of levels $r$, in contrast to the nonuniform condition which is independent of $r$.  We discuss this point further in \S \ref{ss:further}.

Theorem \ref{t:RIPlevels} is in fact a particular instance of the following more general result:

\thm{
[Subsampled isometries and the RIPL II]
\label{t:RIPlevels2}
Let $U \in \bbC^{N \times N}$ be an isometry, $r \in \bbN$, $0 < \epsilon,\delta < 1$ and $0 \leq r_0 \leq r$.  Let $\Omega = \Omega_{\mb{N},\mb{m}}$ be an $(\mb{N},\mb{m})$-multilevel subsampling scheme, and $\mb{M}$ and $\mb{s}$ be sparsity levels and local sparsities respectively.  Suppose that
\bes{
m_k = N_k - N_{k-1},\quad k=1,\ldots,r_0,
}
and
\bes{
m_k \gtrsim \delta^{-2} \cdot (N_k - N_{k-1}) \cdot \left ( \sum^{r}_{l=1} \mu_{k,l} \cdot s_l \right ) \cdot \left ( r \log(2 \tilde{m}) \log(2N) \log^2(2s) + \log(\epsilon^{-1}) \right )
}
for $k=r_0+1,\ldots,r$, where $\tilde{m} = m_{r_0+1} + \ldots + m_r$.  Then with probability at least $1-\epsilon$ the matrix \R{A_multilevel} satisfies the RIPL of order $(\mb{s},\mb{M})$ with constant $\delta_{\mb{s},\mb{M}} \leq \delta$.

}

We shall discuss the significance of this result in \S \ref{ss:whysuperior}.

\subsection{Uniform recovery for 1D Fourier sampling with Haar wavelet sparsity}
\label{ss:unif_FourHaar}
Consider the setup introduced in \S \ref{ss:FourHaar}.  The following result is a uniform counterpart to the nonuniform guarantee Corollary \ref{c:nonunif_FourHaar}:

\cor{
[RIPL for the 1D Fourier/Haar system I]
\label{c:FourierHaar}
Let $N = 2^r$ for some $r \geq 1$ and $0 < \epsilon,\delta < 1$.  For each $k=1,\ldots,r$ suppose that $m_k$ Fourier samples are drawn randomly from the band $W_k$, where
\be{
\label{unif_mk_Haar}
m_k \gtrsim \delta^{-2} \cdot \left ( s_k + \sum^{r}_{\substack{l =1 \\ l \neq k}} 2^{-|k-l|} s_l \right ) \cdot \left ( \log(2m) \log^2(2N) \log^2(2s) + \log(\epsilon^{-1}) \right ) ,\quad k=1,\ldots,r,
}
for some $s_k \leq M_k - M_{k-1}$, $k=1,\ldots,r$, the $M_k$ are as in \R{FourWav_M}, $s = s_1+\ldots+s_r$ and $m=m_1+\ldots+m_r$.  Then with probability at least $1-\epsilon$ the scaled matrix \R{A_multilevel} satisfies the RIPL with constant $\delta_{\mb{s},\mb{M}} \leq \delta$.
}
\prf{
In \cite[Lem.\ 1]{DiscreteFourierHaar} it was proved that $\mu_{k,l} \lesssim 2^{-k} 2^{-|k-l|}$.  Since $N_k - N_{k-1} \leq 2^k$ (see \R{FourWav_N}) we have
\bes{
(N_k - N_{k-1}) \cdot \left ( \sum^{r}_{l=1} \mu_{k,l} \cdot s_l \right ) \lesssim \sum^{r}_{l=1} 2^{-|k-l|} s_{l}.
}
We now apply Theorem \ref{t:RIPlevels}, noting that $r = \log_2(N) \lesssim \log(2N)$.
}

Upon comparison with the nonuniform result Corollary \ref{c:nonunif_FourHaar}, we see that the effect of the interference between separate sparsity levels is substantially smaller in the uniform case.  Namely, the dependence on $s_l$ in \R{unif_mk_Haar} scales like $2^{-|k-l|}$ as opposed to $2^{-|k-l|/2}$ in \R{nonunif_mk_Haar}.

\cor{
[RIPL for the 1D Fourier/Haar system II]
\label{c:FourierHaar2}
Let $N = 2^r$ for some $r \geq 1$ and $0 < \epsilon,\delta < 1$.  For some $0 \leq r_0 \leq r$ suppose that the first $r_0$ frequency bands $W_1,\ldots,W_{r_0}$ are fully sampled, i.e.
\bes{
m_k = N_k - N_{k-1} = 2^{\max\{k-1,1\}},\quad k=1,\ldots,r_0,
}
and  suppose that for $k=r_{0}+1,\ldots,r$, $m_k$ Fourier samples are drawn randomly from the band $W_k$, where
\bes{
m_k \gtrsim \delta^{-2} \cdot \left ( s_k + \sum^{r}_{\substack{l =r_0+1 \\ l \neq k}} 2^{-|k-l|} s_l \right ) \cdot \left ( \log(2 \tilde{m}) \log^2(2N) \log^2(2s) + \log(\epsilon^{-1}) \right ) ,
}
for some $s_k \leq M_k - M_{k-1}$, $k=r_0+1,\ldots,r$, where the $M_k$ are as in \R{FourWav_M}, $s = s_1+\ldots+s_r$ and $\tilde{m} = m_{r_0+1}+\ldots + m_r$.  If $N_{r_0} \leq s_{r_0+1}$ then with probability at least $1-\epsilon$ the scaled matrix \R{A_multilevel} satisfies the RIPL with constant $\delta_{\mb{s},\mb{M}} \leq \delta$, where $\mb{s} = (s_1,\ldots,s_r)$ and $s_k = N_k - N_{k-1} = 2^{\max\{k-1,1\} }$ for $k=1,\ldots,r_0$.
}
\prf{
We apply Theorem \ref{t:RIPlevels2}.  Due to the estimates $N_{k} - N_{k-1} \leq 2^{k}$, $\mu_{k,l} \lesssim 2^{-k} 2^{-|k-l|}$ and the fact that $k > r_0$, we have
\eas{
(N_k-  N_{k-1}) \cdot \left ( \sum^{r}_{l=1} \mu_{k,l} \cdot s_l \right ) & \lesssim  \sum^{r}_{l=1} 2^{-|k-l|} \cdot s_l 
\\
& \lesssim \sum^{r_0}_{l=1} 2^{2l-k} + \sum^{r}_{l=r_0+1} 2^{-|k-l|} \cdot s_l 
\\
& \lesssim 2^{2 r_0 - k} +  \sum^{r}_{l=r_0+1} 2^{-|k-l|} \cdot s_l 
\\
& \leq s_{r_0+1} 2^{r_0-k} +  \sum^{r}_{l=r_0+1} 2^{-|k-l|} \cdot s_l 
\\
& \lesssim \sum^{r}_{l=r_0+1} 2^{-|k-l|} \cdot s_l ,
}
where in the penultimate step we use the fact that $2^{r_0} = N_{r_0} \leq s_{r_0+1}$.
}

\subsection{Why Fourier sampling with wavelets is superior to incoherent sampling}\label{ss:whysuperior}

As mentioned in \S \ref{s:introduction}, when wavelets are used as the sparsifying transform one can obtain superior reconstructions in practice using multilevel subsampled Fourier measurements than with classical incoherent sampling (e.g.\ random Gaussian measurements).  Corollary \ref{c:FourierHaar2} gives some theoretical justification for this empirical observation.  To see wny, we first recall that the number random Gaussian measurements sufficient for the standard RIP is
\be{
\label{randGauss}
m \gtrsim \delta^{-2} \cdot s \cdot \left ( \log(\E N/s) + \log(\epsilon^{-1}) \right ),
}
(this, of course, implies the RIPL)\footnote{We note in passing that the slightly weaker condition $m \gtrsim \delta^{-2} \left ( \sum^{r}_{k=1} s_k \log \left ( \frac{\E (M_k - M_{k-1})}{s_k} \right ) + \log(\epsilon^{-1}) \right )$ suffices for an $m \times N$ Gaussian random matrix to have the RIPL.  This follows from \cite[Cor.\ 5.4]{DirksenDimensionality}, noting that the $\Sigma_{\mb{s},\mb{M}}$ is a union of $ \prod^{r}_{k=1} \left ( \begin{array}{c} M_{k} - M_{k-1} \\ s_k \end{array} \right ) \leq \prod^{r}_{k=1} \left ( \frac{\E (M_k-M_{k-1})}{s_k} \right )^{s_k}$ subspaces of dimension at most $s$.  However, this bound does not change the conclusion of the ensuing discussion significantly.}.  By means of comparison, if we sum \R{unif_mk_Haar} over $k$ we obtain
\be{
\label{wavelet_energy_alt}
m = m_1+\ldots+m_r \gtrsim \delta^{-2} \cdot s\times \mbox{log factors},
}
which has the same scaling in terms of $s$ as \R{randGauss}, and turns out to be slightly worse in terms of the log factors.  However, now suppose the first $r_0$ levels are fully sampled.  Then, applying Corollary \ref{c:FourierHaar2} instead we arrive at the condition
\be{
\label{wavelet_energy}
m \geq N_{r_0} +C  \cdot \delta^{-2} \cdot \left ( s_{r_0+1} + \ldots + s_r \right ) \times \mbox{log factors},
}
where $C$ is a constant.  Recall that wavelet coefficients tend not to be sparse in the first few, i.e.\ $r_0$, coarsest scales, yet most of the energy of the signal is contained in these coefficients.  The guarantee \R{wavelet_energy} shows that the nonsparse portion of the wavelet coefficients can be recovered by Fourier sampling using an optimal number of measurements equal to exactly $N_{r_0}$; in particular, free from universal constants and possible log factors.  In comparison, Gaussian random measurements (or similar) cannot recover the majority of the signal's energy anywhere near as efficiently.\footnote{It is worth noting that in the case of \R{randGauss} the requirement on $\delta$ for stable and robust recovery is $\delta \lesssim 1$ (recall Theorem \ref{t:RIP_recov}).  Conversely, in  \R{wavelet_energy_alt} and \R{wavelet_energy} the requirement is $\delta \lesssim 1/\sqrt{r \rho}$ (see Theorem \ref{t:RIPLstabrobust}), i.e.\ $\delta^{-2} \gtrsim r \rho$.  Since $N = 2^r$, the $r$ term contributes an additional $\log(N)$ factor to the measurement condition.  The $\rho$ term, equal to $\max_{k,l=1,\ldots,r} \{ s_k/s_l \}$, could potentially blow up depending on the local sparsities $s_k$.  However, as discussed in \S \ref{ss:RIPL}, this term can be removed by considering the weighted $\ell^1$ minimization functional \R{Remiregularizer}.} 

It is for this reason that multilevel subsampling of Fourier matrices substantially outperforms random Gaussian sampling in practice for wavelet-based signal and image reconstruction.  Note that a similar observation has also been made in \cite{ModelsCS}.  We refer also to \cite{AsymptoticCS} for a thorough empirical comparison between incoherent sampling and structured multilevel Fourier subsampling.

\subsection{Further discussion}\label{ss:further}
In \S \ref{ss:unif_FourHaar} we have seen an example where, up to log factors, the uniform recovery guarantees implied by Theorem \ref{t:RIPlevels} is strictly smaller than the nonuniform guarantee of Theorem \ref{t:nonunifom_sparsitylevels}.  However, this need not always be the case, as we now discuss.

To this end, let $N = r n$ for some $n,r \geq 2$ and $U$ be a block diagonal matrix $U = \diag ( U^{(1)},\ldots,U^{(r)})$, where for each $k$, the $k^{\rth}$ diagonal block $U^{(k)} \in \bbC^{n \times n}$ is an isometry.  For simplicity, we assume that each $U^{(k)}$ is incoherent, i.e. $n \mu(U^{(k)}) \lesssim 1$.  It is a straightforward exercise to show that the measurement condition \R{nonunif_mk_1} of the nonuniform recovery guarantee (Theorem \ref{t:nonunifom_sparsitylevels}) reduces to
\be{
\label{BD_nonunif}
m_{k} \gtrsim s_{k} \cdot \log(s \epsilon^{-1}) \cdot \left ( \log(r) + \log(n) \right ),\qquad k=1,\ldots,r.
}
On the other hand, the measurement condition \R{mk_RIPL} in the RIPL guarantee (Theorem \ref{t:RIPlevels}) is
\be{
\label{BD_unif}
m_{k} \gtrsim \delta^{-2} \cdot s_k \cdot \left ( r \log(m) \left ( \log(r) + \log(n) \right ) \log^2(s) + \log(\epsilon^{-1}) \right ),\qquad k=1,\ldots,r.
}
Of course, when $U$ is block diagonal the problem of recovering $x \in \bbC^{N}$ decouples into $r$ subproblems of recovering the vectors $x^{(k)} = P^{N_{k-1}}_{N_k} x$ from measurements $y^{(k)} = P_{\Omega_k} U^{(k)} x^{(k)}$.  As discussed in \cite{Candes_Plan}, coherence provides an informatic-theoretic limit, meaning that the minimum number of measurements $m_k$ in the $k^{\rth}$ subproblem has to be of order
\bes{
s_k \cdot \log(n).
}
Hence the nonuniform measurement condition \R{BD_nonunif} is optimal, up to the logarithmic term in the failure probability $\log(s \epsilon^{-1})$, provided $r \lesssim n$.

Conversely, even under the assumption $r \lesssim n$, the uniform measurement condition \R{BD_unif} fails to be optimal.  First, there is additional logarithmic factor $\log(m) \log^2(s)$.  This, however, is not an artefact of the sparsity in levels model per se, but rather a factor identical to one well-known in classical RIP guarantees for subsampled isometries.  See \S \ref{s:proof} for some further discussion.  On the other hand, \R{BD_unif} also exhibits a linear dependence on $r$ that is not seen in \R{BD_nonunif}.  And moreover, while $\delta^{-2} \gtrsim 1$ is sufficient for the classical RIP (Theorem \ref{t:RIP_recov}), for the RIPL one requires $\delta^{-2} \gtrsim \rho r$, where $\rho$ is the sparsity ratio \R{sparsity_ratio} (see Theorem \ref{t:RIPLstabrobust}).  Note that $\rho$ could be removed via the modified regularizer \R{Remiregularizer}, but that still leaves an overall quadratic dependence on the number of levels $r$ in the measurement condition.  Naturally, this scaling is not sharp for the above problem.  The separability implies that the measurement condition required for uniform recovery of each $x^{(k)}$ is no more than $m_{k} \gtrsim s_{k} \cdot (\log(m) \log(n) \log^2(s)+\log(\epsilon^{-1}))$, i.e.\ independent of $r$.

This discussion raises the question of whether the scaling in $r$ can be removed.  We offer some further comments on this topic in \S \ref{s:conclusions}.

\section{Proof of Theorem \ref{t:RIPlevels2}}\label{s:proof}
The proof of Theorem \ref{t:RIPlevels2} is a generalization of that formulated in \cite{RauhutLinz} for the standard RIP for so-called bounded orthonormal systems.  See \cite{FoucartRauhutCSbook} for a historical overview.  The general strategy is as follows.  First, we use Dudley's inequality to bound the expectation of the RICL $\delta_{\mb{s},\mb{M}}$ in terms of an integral of the covering numbers of the set of unit $(\mb{s},\mb{M})$-sparse vectors.  These covering numbers are estimated for small and large values of the covering radius $t$ (the latter via Maurey's method), leading to an overall bound for $\bbE(\delta_{\mb{s},\mb{M}})$.  In the second step, we use a deviation inequality to bound the probability of $\delta_{\mb{s},\mb{M}}$ differing from its expectation.  Combined with the first step, this results in the required probabilistic bound for the event $\delta_{\mb{s},\mb{M}} \leq \delta$.

Note that there have been a number of subsequent developments in proving the classical RIP for bounded orthonormal systems, leading to somewhat smaller log factors in the recovery guarantees \cite{AnderssonStrombergRIP,CheraghchiEtAlSODA,RauhutWardWeighted}.  Due to the additional complications of the sparsity in levels signal model, we have so far not found these approaches to be successful in proving RIPL results.  We also note the recent work of \cite{ChkifaDownwardsCS} in which an improvement of one log factor in $s$ is obtained, at the expense of a worse scaling in $\delta$.  Since $\delta^{-2}$ in our case is required to behave like $r \rho$, where $r$ is the number of levels and $\rho$ is the sparsity ratio (recall Theorem \ref{t:RIPLstabrobust}), we opt not to follow this approach.

\subsection{Setup}
Let $U \in \bbC^{N \times N}$ be an isometry.  Define the sampling levels $\mb{N} = (N_1,\ldots,N_r)$, sparsity levels $\mb{M} = (M_1,\ldots,M_r)$, sparsities $\mb{s} = (s_1,\ldots,s_r)$ and local numbers of samples $\mb{m} = (m_1,\ldots,m_r)$, where $N_r = M_r = N$ and $m_{k} = N_k - N_{k-1}$ for $k=1,\ldots,r_0$.  For $k=1,\ldots,r_0$, let $t_{k,i} = N_{k-1}+i$, $i=1,\ldots,m_k$.  Conversely, for each $k=r_0+1,\ldots,r$, draw $m_{k}$ numbers $t_{k,1},\ldots,t_{k,m_k}$ independently and uniformly at random from the set $\{ N_{k-1}+1,\ldots,N_k\}$.  Note that such numbers need not be distinct.
Let
\bes{
\Omega_{k} = \{ t_{k,1},\ldots,t_{k,m_k} \},\qquad k=1,\ldots,r,
}
and write $\Omega = \Omega_1 \cup \cdots \cup \Omega_r$ for the corresponding $(\mb{m},\mb{N})$-multilevel random subsampling pattern.  Consider the rescaled matrix
\bes{
A = \left [ \begin{array}{c} 1/\sqrt{p_1} P_{\Omega_1} U \\ 1/\sqrt{p_2} P_{\Omega_2} U \\ \vdots \\ 1/\sqrt{p_r} P_{\Omega_r} U \end{array} \right ] \in \bbC^{m\times N},\qquad m = m_1+\ldots + m_r,
}
where $p_{k} = \frac{m_k}{N_k-N_{k-1}}$.  Note that $p_k = 1$ and $\Omega_{k} = \{ N_{k-1}+1,\ldots,N_k\}$ for $k=1,\ldots,r_0$.  

We first observe that the matrix $P_{\Omega_k}$ can be written as
\bes{
P_{\Omega_k} = \sum^{m_k}_{i=1} e_{t_{k,i}} e^*_{t_{k,i}},
}
where $\{ e_i \}^{N}_{i=1}$ is the standard basis on $\bbC^N$.  It now follows that
\be{
\label{AstarA_sum}
A^* A = \sum^{r}_{k=1} \frac{1}{p_k} U^* P_{\Omega_k} U = \sum^{r}_{k=1} \frac{1}{p_k} \sum^{m_k}_{i=1} U^*e_{t_{k,i}} e^*_{t_{k,i}} U = U^* P_{N_{r_0}} U + \sum^{r}_{k=r_0+1} \sum^{m_k}_{i=1} X_{k,i} X^*_{k,i},
}
where $X_{k,i}$ are random vectors given by $X_{k,i} = \frac{1}{\sqrt{p_k}} U^* e_{t_{k,i}}$.  Note that the $X_{k,i}$ are independent, and also that
\ea{
\bbE(A^*A) &= U^* P_{N_{r_0}} U + \sum^{r}_{k=r_0+1} \sum^{m_k}_{i=1} \bbE \left ( X_{k,i} X^*_{k,i} \right ) \nn
\\
& = U^* P_{N_{r_0}} U + \sum^{r}_{k=r_0+1} \frac{m_k}{p_k(N_k-N_{k-1})} \sum^{N_k}_{j=N_{k-1}+1} U^* e_j e^*_j U \nn
\\
& = U^* P_{N_{r_0}} U + U^* P^{\perp}_{N_{r_0}} U \nn
\\
& = U^* U = I. \label{AstarA_expectation}
}
Here, in the second equality we use the fact that $t_{k,i}$ is a random variable taking values $j \in \{ N_{k-1}+1,\ldots,N_k\}$ with equal probability $\frac{1}{N_k-N_{k-1}}$.
Let $D_{\mb{s},\mb{M}} $ be the set of $(\mb{s},\mb{M})$-sparse vectors with $\ell^2$-norm at most one.  That is to say
\be{
\label{D_union}
D_{\mb{s},\mb{M}} = \bigcup_{\Delta \in E_{\mb{s},\mb{M}}} B_{\Delta},
}
where
\bes{
E_{\mb{s},\mb{M}} = \left \{ \Delta \subseteq \{1,\ldots,N \} : | \Delta \cap \{ M_{k-1}+1 , \ldots , M_k \} | \leq s_k,\ k=1,\ldots,r \right \},
}
and
\be{
\label{BS_def}
B_{\Delta} = \left \{ \xi \in \bbC^N : \nm{\xi} \leq 1,\ \supp(\xi) \subseteq \Delta \right \},\qquad \Delta \subseteq \{1,\ldots,N\}.
}
We now define the following seminorm on $\bbC^{N \times N}$:
\bes
{
\tnm{B}_{\mb{s},\mb{M}} := \sup_{z \in D_{\mb{s},\mb{M}}} \left| \ip{B z}{z} \right|.
}
It follows that the restricted isometry constant in levels $\delta_{\mb{s},\mb{M}}$ satisfies
\bes{
\delta_{\mb{s},\mb{M}} = \tnm{A^* A - I }_{\mb{s},\mb{M}},
}
and due to \R{AstarA_sum} and \R{AstarA_expectation} we may rewrite this as
\be{
\label{delta_norm_equiv}
\delta_{\mb{s},\mb{M}}= \tnm{\sum^{r}_{k=r_{0}+1} \sum^{m_k}_{i=1} \left ( X_{k,i} X^*_{k,i} - \bbE (X_{k,i} X^*_{k,i}) \right ) }_{\mb{s},\mb{M}}.
}

\subsection{Estimation of $\bbE (\delta_{\mb{s},\mb{M}})$}
By symmetrization (see \cite[Lem.\ 8.4]{FoucartRauhutCSbook}), we have
\bes{
\bbE \left ( \delta_{\mb{s},\mb{M}} \right ) \leq 2 \bbE\tnm{\sum^{r}_{k=r_{0}+1} \sum^{m_k}_{i=1} \epsilon_{k,i} X_{k,i} X^*_{k,i} }_{\mb{s},\mb{M}},
}
where $\{\epsilon_{k,i} : i=1, \ldots, m_k, k=r_0+1, \ldots, r \}$ is a Rademacher sequence (i.e.\ a sequence of independent random variables each taking the values $+1$ and $-1$ with equal probability) independent of the sampling points $\{ t_{k,i} : i=1, \ldots, m_k, k=r_0+1, \ldots, r \}$.  By definition
\bes{
\bbE_{\epsilon} \tnm{\sum^{r}_{k=r_{0}+1} \sum^{m_k}_{i=1} \epsilon_{k,i} X_{k,i} X^*_{k,i} }_{\mb{s},\mb{M}} = \bbE_{\epsilon} \sup_{u \in D_{\mb{s},\mb{M}}} \left | \sum^{r}_{k=r_{0}+1} \sum^{m_k}_{i=1} \epsilon_{k,i} | \ip{X_{k,i}}{u} |^2 \right |.
}
Conditional on the $X_{k,i}$, this is the supremum of the absolute value of the Rademacher process
\be{
\label{Rademacher_process}
Z_{u} = \sum^{r}_{k=r_{0}+1} \sum^{m_k}_{i=1} \epsilon_{k,i} | \ip{X_{k,i}}{u} |^2,\qquad u \in D_{\mb{s},\mb{M}},
}
with corresponding pseudometric
\bes{
d(u,v) = \left ( \sum^{r}_{k=r_{0}+1} \sum^{m_k}_{i=1} \left ( | \ip{X_{k,i}}{u} |^2 - | \ip{X_{k,i}}{v} |^2 \right )^2 \right )^{1/2},\qquad u,v \in \bbC^N.
}
Observe that
\eas{
\left(| \ip{X_{k,i}}{u} |^2 - | \ip{X_{k,i}}{v} |^2 \right )^2 &= \left ( | \ip{X_{k,i}}{u} | - | \ip{X_{k,i}}{v} | \right )^2 \left ( | \ip{X_{k,i}}{u} | + | \ip{X_{k,i}}{v} | \right )^2 
\\
& \leq | \ip{X_{k,i}}{u-v} |^2  \left ( | \ip{X_{k,i}}{u} | + | \ip{X_{k,i}}{v} | \right )^2
\\
& \leq 2 | \ip{X_{k,i}}{u-v} |^2 \left ( | \ip{X_{k,i}}{u} |^2 + | \ip{X_{k,i}}{v} |^2 \right ),
}
and therefore
\bes{
d(u,v) \leq 2 R \max_{k=r_{0}+1,\ldots,r} \max_{i=1, \ldots, m_{k}}| \ip{X_{k,i}}{u-v} |,
}
where
\be{
\label{Rdef}
R = \sup_{z \in D_{\mb{s},\mb{M}}} \sqrt{\sum^{r}_{k=r_{0}+1} \sum^{m_k}_{i=1} | \ip{X_{k,i}}{z} |^2 } = \sqrt{\tnm{\sum^{r}_{k=r_{0}+1} \sum^{m_k}_{i=1} X_{k,i} X^*_{k,i} }_{\mb{s},\mb{M}}}.
}
If we define the seminorm
\bes{
\| u \|_{X} = \max_{k=r_{0}+1,\ldots,r} \max_{i=1, \ldots, m_{k}}| \ip{X_{k,i}}{u} |,\qquad u \in \bbC^N,
}
then
\be{
\label{pseudometric_bound}
d(u,v) \leq 2 R \nm{u-v}_{X},
}
and we see that the rescaled process $Z_u / (2R)$ satisfies
\bes{
\left ( \bbE \left | Z_u / (2R) - Z_v / (2R) \right |^2 \right )^{1/2} \leq \nm{u-v}_{X}.
}
Dudley's inequality \cite[Thm.\ 8.23]{FoucartRauhutCSbook} now gives
\eas{
\bbE_{\epsilon} \tnm{\sum^{r}_{k=r_{0}+1} \sum^{m_k}_{i=1} \epsilon_{k,i} X_{k,i} X^*_{k,i} }_{\mb{s},\mb{M}} & = 2R \bbE_{\epsilon} \sup_{u \in D_{\mb{s},\mb{M}}} \left | Z_u / (2R ) \right |
\\
& \leq 8 \sqrt{2} R \int^{\Delta (D_{\mb{s}, \mb{M}})/2}_{0} \sqrt{\log(2\cN(D_{\mb{s},\mb{M}} , d/(2R) , t ) )} \D t
\\
& \leq 8 \sqrt{2} R \int^{\Delta (D_{\mb{s}, \mb{M}})/2}_{0} \sqrt{\log(2\cN(D_{\mb{s},\mb{M}} , \nm{\cdot}_X , t ) )} \D t
}
where
\be{
\label{D_radius}
\Delta(D_{\mb{s},\mb{M}}) = \frac{1}{2R} \sup_{u \in D_{\mb{s},\mb{M}}} \sqrt{\bbE | Z_u |^2},
}
and $Z_u$ is as in \R{Rademacher_process}.  Here, in the second inequality we use the fact that $\cN(T,d,t) \leq \cN(T,d',t)$ whenever $d(u,v) \leq d'(u,v)$, $\forall u,v \in T$ (see \cite[\S C.2]{FoucartRauhutCSbook}).  Observe that $\Delta(D_{\mb{s},\mb{M}}) \leq \sup \limits_{u \in D_{\mb{s},\mb{M}}} \| u \|_X$ by \R{pseudometric_bound}.   Also, by definition of the $X_{k,i}$, we have
\ea{
\| u \|_{X} &= \max_{k=r_{0}+1,\ldots,r} \max_{i=1, \ldots, m_{k}} \frac{1}{\sqrt{p_k}} | \ip{U^* e_{t_{k,i}}}{u} | \nn
\\
& \leq \max_{k=r_{0}+1,\ldots,r} \max_{i=1, \ldots, m_{k}} \frac{1}{\sqrt{p_k}} \sum^{r}_{l=1} \sum^{M_l}_{j=M_{l-1}} | U_{t_{k,i},j} | | u_j | \nn
\\
& \leq \max_{k=r_{0}+1,\ldots,r}  \frac{1}{\sqrt{p_k}} \sum^{r}_{l=1} \sqrt{\mu_{k,l}} \sqrt{s_l} \sqrt{\sum^{M_l}_{j=M_{l-1}} | u_j |^2} \nn
\\
& \leq \max_{k=r_{0}+1,\ldots,r} \sqrt{\sum^{r}_{k=1} \frac{\mu_{k,l} s_l}{p_k} } \| u \| = \sqrt{Q} \| u \|, \label{Xnorm_bound}
}
(note that the fourth line follows from the Cauchy--Schwarz inequality), where $Q$ is defined by
\be{
\label{Q_def}
Q = \max_{k=r_0+1,\ldots,r} \sum^{r}_{l=1} \frac{\mu_{k,l} s_l}{p_k}.
}
Since $\| u \| \leq 1$ for $u \in D_{\mb{s},\mb{M}}$ we get $\Delta(D_{\mb{s},\mb{M}}) \leq \sqrt{Q}$, and therefore
\be{
\label{dudley1}
\bbE_{\epsilon} \tnm{\sum^{r}_{k=r_{0}+1} \sum^{m_k}_{i=1} \epsilon_{k,i} X_{k,i} X^*_{k,i} }_{\mb{s},\mb{M}} \leq 8 \sqrt{2} R \int^{\sqrt{Q}/2}_{0} \sqrt{\log(2\cN(D_{\mb{s},\mb{M}} , \nm{\cdot}_X , t ) )} \D t.
}
To estimate this integral we next bound the covering number in the small $t$ and large $t$ regimes respectively.

\subsubsection{Estimate of the covering number $\cN(D_{\mb{s},\mb{M}} , \nm{\cdot}_X , t )$ for small $t$}
Recalling \R{D_union} and standard properties of covering numbers, we notice that
\bes{
\cN(D_{\mb{s},\mb{M}} , \nm{\cdot}_X , t ) \leq \sum_{\Delta \in E_{\mb{s},\mb{M}}} \cN(B_\Delta , \nm{\cdot}_{X} , t ),
}
where $B_\Delta$ is as in \R{BS_def}.  Using the bound \R{Xnorm_bound} and several further properties of covering numbers, we find that
\bes{
\cN(D_{\mb{s},\mb{M}} , \nm{\cdot}_X , t ) \leq \sum_{\Delta \in E_{\mb{s},\mb{M}}} \cN(B_\Delta , \nm{\cdot} , t / \sqrt{Q} ) \leq \sum_{\Delta \in E_{\mb{s},\mb{M}}} \left ( 1 + 2 \sqrt{Q} / t \right )^{2|\Delta|}.
}
Note that $|\Delta| \leq s$ and $|E_{\mb{s},\mb{M}} | \leq \left ( \begin{array}{c} N \\ s \end{array} \right )$, where $s= s_1+\ldots + s_r$.  Hence
\bes{
\cN(D_{\mb{s},\mb{M}} , \nm{\cdot}_X , t ) \leq \left ( \begin{array}{c} N \\ s \end{array} \right ) \left ( 1 + 2 \sqrt{Q} / t \right )^{2s} \leq \left ( \exp(1) N / s \right )^{s} \left ( 1 + 2 \sqrt{Q} / t \right )^{2s},
}
which gives the bound
\be{
\label{cover_number_bound1}
\sqrt{\log \left ( 2 \cN(D_{\mb{s},\mb{M}} , \nm{\cdot}_X , t )  \right )} \leq \sqrt{2s} \left ( \sqrt{\log(2 \exp(1) N / s)} + \sqrt{\log\left ( 1 + 2 \sqrt{Q} / t \right ) } \right ),\quad t>0.
}

\subsubsection{Estimate of the covering number $\cN(D_{\mb{s},\mb{M}} , \nm{\cdot}_X , t )$ for large $t$}\label{ss:larget}

We use Maurey's method.  Let $x \in D_{\mb{s},\mb{M}}$ and write $x = x^{(1)} + \ldots + x^{(r)}$, where
\bes{
\supp(x_k) \subseteq \{ M_{k-1}+1,\ldots,M_k \},\qquad \| x^{(k)} \|_0 \leq s_k.
}
Note that $\| x \|^2 = \sum^{r}_{k=1} \| x^{(k)} \|^2$ and in particular, $\| x^{(k)} \| \leq 1$ for each $k=1,\ldots,r$.  Let $\nm{\cdot}^*_1$ be defined by
\bes{
\nm{z}^*_1 = \sum^{N}_{i=1} \left ( | \Re(z_i) | + | \Im(z_i) | \right ),\qquad z \in \bbC^N,
}
which is the usual $\ell^1$-norm after identifying $\bbC^N$ with $\bbR^{2N}$.  Observe that $\| x^{(k)} \|^*_{1} \leq \sqrt{2s_k}$.  Hence we may write $x^{(k)}$ as the convex combination
\bes{
x^{(k)} = \sum^{4(M_k-M_{k-1})}_{j=1} \lambda^{(k)}_j v^{(k)}_j,
}
where the vectors $v^{(k)}_j$ are an enumeration of the $4(M_k-M_{k-1})$ vectors
\bes{
+ \sqrt{2s_k} e_i , - \sqrt{2s_k} e_i, + \I \sqrt{2s_k} e_i, - \I \sqrt{2s_k} e_i,\qquad i=M_{k-1}+1,\ldots,M_k.
}
Let $Z^{(1)},\ldots,Z^{(r)}$ be independent random vectors such that $Z^{(k)}$ takes value $v^{(k)}_j$ with probability $\lambda^{(k)}_j$.  Since $\lambda^{(k)}_j \geq 0$ and $\sum^{4(M_k-M_{k-1})}_{j=1} \lambda^{(k)}_j = 1$ this defines a valid probability distribution for each $k$.  Note that
\bes{
\bbE \sum^{r}_{k=1} Z^{(k)} = \sum^{r}_{k=1} \sum^{4(M_k-M_{k-1})}_{j=1} \lambda^{(k)}_j v^{(k)}_j = \sum^{r}_{k=1} x^{(k)} = x.
}
Let $R_1,\ldots,R_r \in \bbN$ be integers whose values will be fixed later.  For each $k$, let $Z^{(k)}_{1},\ldots,Z^{(k)}_{R_k}$ be independent copies of $Z^{(k)}$, and consider the sum
\bes{
z = \sum^{r}_{k=1} \frac{1}{R_k} \sum^{R_k}_{q=1} Z^{(k)}_{q}.
}
By symmetrization, we now have
\eas{
\bbE \| z - x \|_{X} &= \bbE\nm{\sum^{r}_{k=1} \frac{1}{R_k} \sum^{R_k}_{q=1} \left ( Z^{(k)}_q - \bbE Z^{(k)}_q \right ) }_{X}
\\
& \leq 2 \bbE\nm{\sum^{r}_{k=1} \frac{1}{R_k} \sum^{R_k}_{q=1} \epsilon^{(k)}_q Z^{(k)}_q }_{X}
\\
& \leq 2 \bbE \max_{l=r_{0}+1,\ldots,r} \max_{i=1, \ldots, m_l} \left | \sum^{r}_{k=1} \frac{1}{R_k} \sum^{R_k}_{q=1} \epsilon^{(k)}_q  \ip{X_{l,i}}{Z^{(k)}_q} \right |,
}
where $\{ \epsilon^{(k)}_{q} : q=1,\ldots,R_k, k=1,\ldots,r \}$ is a Rademacher sequence independent of the $Z^{(k)}_{q}$.  Fix a realization of the $ Z^{(k)}_{q}$.  Recall that $X_{l,i} = \frac{1}{\sqrt{p_l}} U^* e_{t_{l,i}}$ and that each $Z^{(k)}_{q}$ is equal to $v^{(k)}_j$ for some $j=1,\ldots,4(M_k-M_{k-1})$.  Therefore
\bes{
\sum^{r}_{k=1}  \frac{1}{R^2_k}  \sum^{R_k}_{q=1}| \ip{X_{l,i}}{Z^{(k)}_q} |^2 \leq \sum^{r}_{k=1} \frac{1}{R^2_k} \sum^{R_k}_{q=1} \frac{2 s_k \mu_{l,k}}{p_l} = \frac{2}{p_l} \sum^{r}_{k=1} \frac{s_k \mu_{l,k}}{R_k} ,
}
from which it follows that
\bes{
\left ( \sum^{r}_{k=1}  \frac{1}{R^2_k}  \sum^{R_k}_{q=1}| \ip{X_i}{Z^{(k)}_q} |^2 \right )^{1/2} \leq \sqrt{A},\qquad A = \max_{l=r_{0}+1,\ldots,r} \sum^{r}_{k=1} \frac{2s_k \mu_{l,k}}{p_l R_k}.
}
We can now define the random variable $Y_{l,i} := \sum \limits_{k=1}^r \frac{1}{R_k} \sum \limits_{q=1}^{R_k} \epsilon_q^{(k)} \ip{X_{l,i}}{Z^{(k)}_q} $.  Using \cite[Thm.\ 8.8]{FoucartRauhutCSbook} we deduce that $Y_i$ satisfies
\bes{
\bbP_{\epsilon} \left( |Y_{l,i}|\geq\sqrt{A} y \right) \leq 2 e^{-y^2/2}, \qquad y>0,
}
where $\tilde{m} = \sum \limits^{r}_{k=r_0+1} m_{k}$. By the union bound
\bes{
\bbP_{\epsilon} \left( \max_{l=r_{0}+1,\ldots,r} \max_{i=1, \ldots, m_l} |Y_{l,i}|\geq \sqrt{A} y \right) \leq 2 \tilde{m} e^{-y^2/2},\qquad y>0,
}
and hence \cite[Prop.\ 7.14]{FoucartRauhutCSbook} gives
\bes{
\bbE_{\epsilon}  \max_{l=r_{0}+1, \ldots, r} \max_{i=1,\ldots,m_{l}} \left | \sum^{r}_{k=1} \frac{1}{R_k} \sum^{R_k}_{q=1} \epsilon^{(k)}_q  \ip{X_{l,i}}{Z^{(k)}_t} \right |  \leq 3/2 \sqrt{A} \sqrt{\log(8 \tilde{m})}.
}
Using Fubini's theorem, we now obtain the bound
\bes{
\bbE\| x - z \|_{X} \leq 2 \bbE_Z \bbE_{\epsilon} \max_{l=r_{0}+1, \ldots, r} \max_{i=1,\ldots,m_{l}} \left | \sum^{r}_{k=1} \frac{1}{R_k} \sum^{R_k}_{q=1} \epsilon^{(k)}_t  \ip{X_{l,i}}{Z^{(k)}_q} \right | \leq 3 \sqrt{A} \sqrt{\log(8 \tilde{m})}.
}
This means that there exists a vector $z$ of the form
\bes{
z = \sum_{k=1}^r \frac{1}{R_k}\sum_{q=1}^{R_k} Z^{(k)}_q,
}
with $\| x - z \|_{X} \leq 3 \sqrt{A} \sqrt{\log(8 \tilde{m})}$.  In particular, $\| x - z \|_{X} \leq t / 2$ provided
\be{
\label{A_cond1}
3 \sqrt{A} \sqrt{\log(8 \tilde{m})} \leq t /2.
}
Note that $z$ can take at most $(4N)^{R_1+\ldots+R_r}$ values.  Hence we have constructed a set of $(4N)^{R_1+\ldots+R_r}$ points such that, for any $x \in D_{\mb{s},\mb{M}}$, we have $\| x - z \|_{X} \leq t/2$ for some $z$ in the set.  To construct a cover, we need the points to belong to $D_{\mb{s},\mb{M}}$.  As in the usual way, for each $z$ we can find a point $z' \in D_{\mb{s},\mb{M}}$ with $\| z - z' \|_{X} \leq t/2$.  If no such point exists then we discard that $z$, since it will not be needed in the cover.  After doing this, we then deduce that there exists a cover of size at most $(4N)^{R_1+\ldots+R_r}$ such that for any $x \in D_{\mb{s},\mb{M}}$ we have $\| x - z \|_{X} \leq t$ for some $z$ in the cover.  Thus
\bes{
\log(2 \cN(D_{\mb{s},\mb{M}} , \nm{\cdot}_X , t )) \leq \left ( \sum^{r}_{k=1} R_k \right )\log(8 N),
}
provided $R_1,\ldots,R_r$ are chosen so that \R{A_cond1} holds.  To do this, we set
\bes{
R_1 = \ldots = R_r = \left \lceil \frac{36}{t^2} \log(8 \tilde{m}) \max_{l =r_{0}+1,\ldots,r} \sum^{r}_{k=1} \frac{2 s_k \mu_{l,k}}{p_l} \right \rceil = \left \lceil \frac{72}{t^2} Q \log(8 \tilde{m}) \right \rceil .
}
Using these values, we obtain the following bound for the covering number:
\ea{
\sqrt{\log(2 \cN(D_{\mb{s},\mb{M}} , \nm{\cdot}_X , t ))} &\leq \sqrt{r \left ( 72 Q \log(8 \tilde{m}) / t^2 +  1 \right ) \log(8N)} \nn
\\
& \leq 12 \sqrt{r} \sqrt{Q} \sqrt{\log(8N)\log(8 \tilde{m})} / t,\quad 0 < t \leq \sqrt{Q}/2. \label{cover_number_bound2}
}
Note that in the last step we use the fact that $t \leq \sqrt{Q}/2$ in \R{dudley1}.

\subsubsection{Overall bound for $\bbE(\delta_{\mb{s},\mb{M}})$}
With \R{cover_number_bound1} and \R{cover_number_bound2} in hand, we return to \R{dudley1}.  Splitting the range of integration into $(0,\tau)$ and $(\tau , \sqrt{Q}/2)$ for some $0 < \tau < \sqrt{Q}/2$ to be determined and using the integral inequality $\int^{\alpha}_{0} \sqrt{\log(1+1/t)} \D t \leq \alpha \sqrt{\log(\exp(1)(1+1/\alpha))}$ \cite[Lem.\ C.9]{FoucartRauhutCSbook}, we find that
\eas{
\bbE_{\epsilon} \tnm{\sum^{r}_{k=r_{0}+1} \sum^{m_k}_{i=1} \epsilon_{k,i} X_{k,i} X^*_{k,i} }_{\mb{s},\mb{M}}
\leq & 8 \sqrt{2} R \Bigg ( \sqrt{2s} \left (\sqrt{\log(2 \exp(1) N / s)} + \sqrt{\log \left ( \exp(1) (1 + 2 \sqrt{Q} / \tau ) \right )} \right ) \tau
\\
& + 12 \sqrt{r} \sqrt{Q} \sqrt{\log(8N) \log(8 \tilde{m})} \log\left(\sqrt{Q} /(2 \tau) \right ) \Bigg ).
}
We now set $\tau = \sqrt{Q} / (3 \sqrt{s})$ to get
\bes{
\bbE_{\epsilon} \tnm{\sum^{r}_{k=r_{0}+1} \sum^{m_k}_{i=1} \epsilon_{k,i} X_{k,i} X^*_{k,i} }_{\mb{s},\mb{M}} \leq C R \sqrt{Q} \sqrt{r} \sqrt{\log(2 \tilde{m}) \log(2N)} \log(2s),
}
for some constant $C>0$.  Hence, by \R{delta_norm_equiv}, \R{Rdef} and Fubini's theorem we arrive at the estimate
\be{
\label{delta_expectation_sqrt}
\bbE \delta_{\mb{s},\mb{M}} \leq C \sqrt{Q} \sqrt{r} \sqrt{\log(2 \tilde{m}) \log(2N)} \log(2s) \bbE \sqrt{\tnm{\sum^{r}_{k=r_{0}+1} \sum^{m_k}_{i=1} X_{k,i} X^*_{k,i} }_{\mb{s},\mb{M}}}.
}
Recall from \R{AstarA_expectation} that
\bes{
\sum^{r}_{k=r_{0}+1} \sum^{m_k}_{i=1} \bbE X_{k,i} X^*_{k,i} = U^* P^{\perp}_{N_{r_0}} U,
}
and therefore
\bes{
\nm{\sum^{r}_{k=r_{0}+1} \sum^{m_k}_{i=1} \bbE X_{k,i} X^*_{k,i} }_{\mb{s},\mb{M}} = \sup_{z \in D_{\mb{s},\mb{M}}} \| P^{\perp}_{N_{r_0}} U z \|^2 \leq 1,
}
since $U$ is an isometry.  Hence, by the Cauchy--Schwarz inequality, we have
\bes{
\bbE \sqrt{\tnm{\sum^{r}_{k=r_{0}+1} \sum^{m_k}_{i=1} X_{k,i} X^*_{k,i} }_{\mb{s},\mb{M}}} \leq \bbE \sqrt{\tnm{\sum^{r}_{k=r_{0}+1} \sum^{m_k}_{i=1} \left (X_{k,i} X^*_{k,i} -\bbE X_{k,i} X^*_{k,i} \right ) }_{\mb{s},\mb{M}} + 1} \leq \sqrt{\bbE  \left ( \delta_{\mb{s},\mb{M}} \right ) + 1}.
}
Combining this with \R{delta_expectation_sqrt} now gives
\be{
\label{delta_expectation}
\bbE \left ( \delta_{\mb{s},\mb{M}} \right ) \leq D + D^2,\qquad D = C \sqrt{Q} \sqrt{r} \sqrt{ \log(2 \tilde{m}) \log(2N)} \log(2s).
}

\subsection{Estimate for $\delta_{\mb{s},\mb{M}}$}
Having estimated its expectation, we are now able to provide a probabilistic bound for $\delta_{\mb{s},\mb{M}}$ itself.  We shall use Theorem 8.42 of \cite{FoucartRauhutCSbook}.  What follows below is a setup to apply this result.  First, \R{AstarA_sum} and \R{AstarA_expectation} give that
\bes{
\delta_{\mb{s},\mb{M}} = \tnm{\sum^{r}_{k=r_0+1} \sum^{m_k}_{i=1} X_{k,i} X^*_{k,i} - U^* P^{\perp}_{N_{r_0}} U }_{\mb{s},\mb{M}} =  \tnm{\sum^{r}_{k=r_0+1} \sum^{m_k}_{i=1} \left ( X_{k,i} X^*_{k,i} - \frac{1}{m_k} U^* P^{N_{k-1}}_{N_k} U \right ) }_{\mb{s},\mb{M}} ,
}
and therefore
\bes{
\delta_{\mb{s},\mb{M}} = \sup_{(z,w) \in Q^*_{\mb{s},\mb{M}}} \Re \ipl{ \sum^{r}_{k=r_0+1} \sum^{m_k}_{i=1} Y_{k,i} w }{z}
}
where $Q^*_{\mb{s},\mb{M}}$ is a countable dense subset of
\bes{
Q_{\mb{s},\mb{M}} = \bigcup_{\Delta \in E_{\mb{s},\mb{M}} } \left \{ (z,w) : \| z \| = \| w \| =1,\  \supp(z) = \supp(w) \subseteq \Delta \right \}.
}
and
\bes{
Y_{k,i} = X_{k,i} X^*_{k,i} - \frac{1}{m_k} U^* P^{N_{k-1}}_{N_k} U.
}
Hence
\be{
\label{TalagrandFun}
\delta_{\mb{s},\mb{M}}   = \sup_{(z,w) \in Q^*_{\mb{s},\mb{M}}} \sum^{r}_{k=r_0+1} \sum^{m_k}_{i=1} f_{z,w}(Y_{k,i}),
}
where $f_{z,w}$ are the functions given by $f_{z,w} : \bbC^{N \times N} \rightarrow \bbR, Y \mapsto \Re \ip{Y w}{z}$.
Observe that $\bbE f_{z,w}(Y_{l,i}) = 0$.  Also we have
\eas{
| f_{z,w}(Y_{k,i}) | & \leq | \ip{ (X_{k,i} X^*_{k,i} - \frac{1}{m_k} U^* P^{N_{k-1}}_{N_k} U) w}{z} |
\\
 &\leq  | \ip{ X_{k,i} X^*_{k,i} w}{z} | + | \ip{ \frac{1}{m_k} U^* P^{N_{k-1}}_{N_k} U w}{z} |
 \\
& \leq \frac{1}{p_k} | \ip{e_{t_{k,i}}}{U w} | \ip{e_{t_{k,i}}}{U z} | + \frac{1}{m_k} \sum^{N_k}_{j=N_{k-1}+1} | \ip{e_j}{U w} | | \ip{e_j}{U z} |
\\
& \leq \frac{2}{p_k} \max_{j = N_{k-1}+1,\ldots,N_k} | \ip{e_j}{U w} | | \ip{e_j}{U z} | \leq 2Q,
}
since
\bes{
| \ip{U^*e_{j}}{z} | \leq \sum^{r}_{l=1} \sum^{M_{l}}_{i=M_{l-1}+1} |U_{ji} | |z_i| \leq \sqrt{\sum^{r}_{l=1} \mu_{k,l} s_l},
}
and likewise for $| \ip{U^*e_{j}}{w} |$.  Furthermore, we have
\eas{
\bbE \sum^{r}_{k=r_0+1} \sum^{m_k}_{i=1} | f_{z,w}(Y_{k,i}) |^2  & \leq \sum^{r}_{k=r_0+1} \sum^{m_k}_{i=1} \bbE \nm{ \left ( X_{k,i} X^*_{k,i} - \frac{1}{m_k} U^* P^{N_{k-1}}_{N_k} U \right ) w }^2 = S_1 - S_2 + S_3,
}
where
\eas{
S_1 &= \sum^{r}_{k=r_0+1} \sum^{m_k}_{i=1} \bbE \| X_{k,i} P_{\Delta} \|^2 | \ip{X_{k,i}}{P_{\Delta} w} |^2
\\
 S_2 &= \sum^{r}_{k=r_0+1} \sum^{m_k}_{i=1}\frac{1}{m_k} 2 \Re \bbE \ip{X_{k,i}}{w} \ip{X_{k,i}}{U^* P^{N_{k-1}}_{N_k} U w }
 \\
S_3 & = \sum^{r}_{k=r_0+1} \sum^{m_k}_{i=1} \frac{1}{m^2_k} \| U^* P^{N_{k-1}}_{N_k} U w \|^2 = \sum^{r}_{k=r_0+1} \frac{1}{m_k} \| P^{N_{k-1}}_{N_k} U w \|^2 ,
}
and $\Delta$ is the support of $w$.  Arguing as above, we note that 
\bes{
\| X_{k,i} P_{\Delta} \|^2 = \sum_{j \in \Delta} \frac{| \ip{e_{t_{k,i}}}{U e_j} |^2}{p_k} \leq \sum^{r}_{k=1} \frac{\mu_{k,l} s_l}{p_k} \leq Q.
}
Hence
\bes{
S_1 \leq Q \sum^{r}_{k=r_{0}+1} \sum^{m_k}_{i=1} \bbE | \ip{X_{k,i}}{P_{\Delta} w} |^2  = Q \sum^{r}_{k=r_0+1} \| P^{N_{k-1}}_{N_k} U w \|^2 = Q \| P^{\perp}_{N_{r_0}} U w \|^2 \leq Q,
}
since $U$ is an isometry.  Also
\bes{
S_2 = 2 \sum^{r}_{k=r_0+1} \sum^{m_k}_{i=1} \frac{1}{m^2_k} \sum^{N_k}_{j=N_{k-1}+1} \ip{e_j}{U w} \ip{e_j}{U U^* P^{N_{k-1}}_{N_k} U w } = 2 S_3.
}
Therefore
\bes{
\bbE \sum^{r}_{k=r_0+1} \sum^{m_k}_{i=1} | f_{z,w}(Y_{k,i}) |^2 \leq Q - S_3 < Q.
}
It now follows from \cite[Thm.\ 8.42]{FoucartRauhutCSbook} (with values $t = \delta/2$, $K = 2 Q$, $\sigma^2 = Q$ and recalling that $0 < \delta < 1$) and \R{TalagrandFun} that
\bes{
\bbP(\delta_{\mb{s},\mb{M}} \geq \delta ) \leq \exp \left ( -3 \delta^2 / (80 Q) \right ),
}
provided
\bes{
\bbE \left ( \delta_{\mb{s},\mb{M}} \right ) \leq \delta / 2.
}
Rearranging and applying \R{delta_expectation}, we deduce that $\delta_{\mb{s},\mb{M}} \leq \delta$ with probability at least $1-\epsilon$, provided
\bes{
Q \delta^{-2} \log(\epsilon^{-1}) \lesssim 1,
}
and
\bes{
Q \delta^{-2} r \log(2 \tilde{m}) \log(2N) \log^2(2s) \lesssim 1.
}
Recalling the definition of $Q$ now completes the proof.

\section{Conclusions}\label{s:conclusions}
In this paper we have established the first uniform recovery guarantee for subsampled isometries with the sparsity in levels class.  This was done by deriving conditions under which the so-called restricted isometry property in levels (RIPL) holds.  Interestingly, the resulting guarantees are simpler and less stringent than existing nonuniform recovery guarantees.  In particular, they do not require relative sparsities (see Definition \ref{d:rel_sparsity}) and involve smaller local coherence factors (see Definition \ref{d:loc_coh}).  For the particular case of Fourier sampling with Haar wavelets, this leads to a noticeably better recovery guarantee.

There are a number of directions for future work.  First, as discussed in \S \ref{ss:further} the quadratic scaling of the measurement condition with the number of levels $r$ is unfortunate.  Indeed, while $r$ contributes only an additional logarithmic factor in the case of wavelets, in other problems it is crucial to have measurement conditions that are independent of $r$.  This is case in applications to multi-sensor acquisition, for example (see \cite{AdcockChunParallel}.) We expect one power of $r$ to be an artefact of the proof given in \S \ref{s:proof}; specifically, it arises only in the estimate of the covering number for large $t$ given in \S \ref{ss:larget}.  Yet, as discussed in \S \ref{ss:RIPL}, the other power of $r$ cannot in general be removed when considering the $\ell^1$ regularizer.  Constructing a regularizer that avoids this factor is an interesting problem for future work.

This aside, in this paper we have only considered sensing operators that arise as isometries on  finite-dimensional vector spaces.  Yet in a series of papers \cite{BAACHGSCS,AHPRBreaking,AsymptoticCS} so-called \textit{infinite-dimensional} CS has been introduced.  Therein the sampling operator $U$ is an isometry on a separable, infinite-dimensional Hilbert space, as opposed to $\bbC^N$.  Nonuniform recovery guarantees for this setup were introduced in \cite{BAACHGSCS}.  Future work will aim to extend the analysis of this paper to this infinite-dimensional setup.  Note that the RIPL was originally formulated in \cite{BastounisHansen} only for finite-dimensional vector spaces.  However, more recent work \cite{TraonmilinGribonvalRIP} has generalized this concept substantially to include possibly infinite-dimensional spaces.

In terms of specific applications, in this paper we have examined the case of discrete, one-dimensional Haar wavelets with Fourier sampling.  Some possible extensions of this include to the multidimensional setting, to other types of wavelets, and to the continuous (as opposed to discrete) case.  For the latter we refer to \cite{AHPRBreaking} for nonuniform recovery guarantees.  Other extensions include to the setting of binary Walsh--Hadamard measurements, which are a practical alternative to multilevel subsampled Fourier measurements in compressive imaging applications \cite{AsymptoticCS}, and to the case of redundant sparsifying transformations.  See \cite{PoonFrames} for nonuniform recovery guarantees in the latter case.  Furthermore, besides wavelet sparsity, the sparsity in levels model has also recently found use in so-called parallel sensing architectures, with nonuniform recovery guarantees being presented in \cite{AdcockChunParallel}.  Future work will look to generalize the results of this paper to this setting.

\section*{Acknowledgements}
Both authors would like to thank Holger Rauhut and Mary Wootters for independently suggesting the problem tackled in this paper.  BA wishes to acknowledge the support of Alfred P. Sloan Foundation and the Natural Sciences and Engineering Research Council of Canada through grant 611675.  CL wishes to acknowledge the support of Professor Weiran Sun and Simon Fraser University during his research visit there from July to October 2015.

\bibliographystyle{abbrv}
\small
\bibliography{UniformRecoveryRefs}

\end{document}